\documentclass{PoS}

\usepackage{bm}% bold math

\title{Novel linear algebraic theory and one-hundred-million-atom quantum material simulations on the K computer}

\ShortTitle{Novel linear algebraic theory and one-hundred-million-atom quantum material simulations ...}

\author{\speaker{Takeo Hoshi}$^{\rm a,b}$,  Tomohiro Sogabe$^{c,b}$, Takafumi Miyata$^{d,b}$, Dongjin Lee$^{d,b}$, Shao-Liang Zhang$^{d,b}$,  Hiroto Imachi$^{e}$, Yoshifumi Kawai$^{a}$, Yohei Akiyama$^{a}$, Keita Yamazaki$^{a}$ and Seiya Yokoyama$^{a}$  \\
        $^{\rm a}$ Department of Applied Mathematics and Physics, Tottori University, Koyama Minami, Tottori 680-8550, Japan \\
        $^{\rm b}$ Japan Science and Technology Agency, Core Research for Evolutional Science and Technology (JST-CREST), 5, Sanbancho, Chiyoda-ku, Tokyo 102-0075, Japan \\
        $^{\rm c}$ Graduate School of Information Science and Technology, Aichi Prefectural University, 1522-3 Ibaragabasama, Kumabari, Nagakute-cho, Aichi-gun, Aichi 480-1198, Japan \\
        $^{\rm d}$ Department of Computational Science and Engineering, Nagoya University, Furo-cho, Chikusa-ku, Nagoya 464-8603, Japan \\
        $^{\rm e}$ Center for Research and Development of Higher Education, The University of Tokyo, Bunkyo-ku, Tokyo 113-8656, Japan \\
        E-mail: \email{hoshi@damp.tottori-u.ac.jp}}

\abstract{
The present paper gives a review of our recent progress and  latest results 
for novel linear-algebraic algorithms and 
its application to large-scale quantum material simulations or electronic structure calculations.
The algorithms are Krylov-subspace (iterative) solvers for generalized shifted linear equations, in the form of $(zS-H) \bm{x} =  \bm{b}$, 
in stead of conventional generalized eigen-value equation.  
The method was implemented in our order-$N$ calculation code ELSES (http://www.elses.jp/) 
with modelled systems based on ab initio calculations. 
The code realized one-hundred-million-atom, or 100-nm-scale, quantum material simulations on the K computer
in a high parallel efficiency with up to all the built-in processor cores.
The present paper also explains several methodological aspects,
such as use of XML files and \lq novice' mode for general users.
A sparse matrix data library in our real problems ( http://www.elses.jp/matrix/ )
was prepared.
Internal eigen-value problem is discussed as a general need from the quantum material simulation. 
The present study is  a interdisciplinary one and is sometimes
called 'Application-Algorithm-Architecture co-design'.
The co-design will play a crucial role in exa-scale scientific computations.
}

\FullConference{2013 International Workshop on Computational Science and Engineering,\\
		14-17 October 2013\\
		National Taiwan University, Taipei, Taiwan}

\begin{document}

\section{Introduction}

Numerical linear algebra with large matrices 
is a common foundation of high-performance scientific computations
and the present paper focuses on
quantum material simulations (electronic structure calculations).
Large scale electronic structure calculations,  with thousand atoms or more, 
are of great importance both in science and technology.
So far, we have developed
linear-algebraic algorithms and 
the code called \lq ELSES' \cite{ELSES-URL} 
for large-scale calculations in nano material studies. 
\cite{TAKAYAMAI-2004-JPSJ, HOSHI-2005-PRB, TAKAYAMA-2006-PRB, IGUCHI-2007-PRL, SHINAOKA-2008-NEGF, YAMAGEN-2008-SCOCG, SOGABE-2008-ETNA, HOSHI-2009-JPCM-AUNW, HOSHI-AUNW-HANDBOOK, YAMASHITA-2011, TENG-2011, SOGABE-2012-GSQMR, HOSHI-mArnoldi, NISHINO-2012-SSI, HOSHI-2013-KEI-BENCH, LEE-2013-INTERIOR-EIGEN, LEE-2013-EASIAM, HOSHI-2013-KEI-BENCH-APPC}

The present paper presents a review of methodologies with latest results.
The present paper is organized as follows; Sec.~2 is devoted to an overview
of our study for methodologies and application. 
Sec.~3 presents novel linear algebraic solvers and 
the benchmark of their application on the K computer.
Several methodologies used with ELSES are reviewed 
for two mode, \lq novice' and  \lq expert'  modes, for general users in Sec.~4
and the use of Extensible Markup Language (XML) for input/output files in Sec.~5.
As our related studies, 
a sparse matrix collection called \lq ELSES matrix library' \cite{ELSES-MATRIX-LIBRARY}
and the concept of numerical \lq engine' \cite{EIGEN-TEST} are explained in Sec.~6. 
In Sec.~7, internal eigen-value problem is discussed,
as a general need from large-scale electronic structure calculations.
In Sec.~8, 
physical (not mathematical) aspects of methodology is focused on,
in particular, modeled (tight-binding-form) electronic structure theory 
based on  {\it ab initio} calculations.
The summary and outlook appear in Sec.~9.  

\section{Overview of methodology and application}

A mathematical foundation of electronic state calculations is 
a generalized eigen-value equation
\begin{eqnarray}
A  \bm{y}_k = \lambda_k B \bm{y}_k.
 \label{EQ-GEV-EQ}
\end{eqnarray}
In the present paper, as in many cases, 
the matrices $A$ and $B$ are $M \times M$ sparse real-symmetric ones and  $B$ is positive definite.
The matrix size $M$ is proportional to the number of atoms 
in the calculated material $N$ ($M \propto N$).
A standard eigen-value equation also appears among cases 
in which the matrix $B$ is reduced to the unit matrix  ($B = I$).
An eigen value $\lambda_k$ is the energy of one electron and usually called \lq eigen level',
while an eigen vector $\bm{y}_k$ represents the shape of wavefunction (electronic 'wave')  for one electron. 
The use of dense-matrix solvers
requires the computational cost is O$(N^3$)
or proportional to $N^3$, 
which will be a bottle neck in large-scale calculations.
Another issue is algorithms suitable to massive parallelism
in modern supercomputers, such as the K computer.

\begin{figure}[htbp] 
\begin{center}
  \includegraphics[width=14cm]{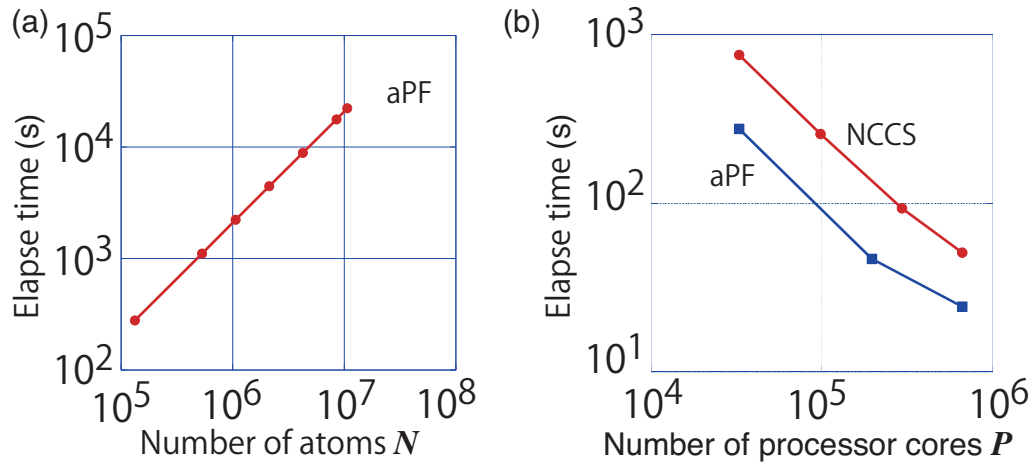}
 %  \vspace{8cm}
\end{center}
%\vspace{-5mm}
\caption{\label{fig-BENCH} 
(a) Benchmark for the order-$N$ scaling. \cite{HOSHI-mArnoldi}
(b) Parallel efficiency on the K computer with one-hundred-million atoms.
(\cite{HOSHI-2013-KEI-BENCH-APPC}, present work). 
The calculations were carried out 
with the number of used processor cores $P$ $(T=T(P))$,
from  $P = P_{\rm min} \equiv 32,768$ to $P_{\rm all} \equiv 663,552$ (all the processor cores).
See the text for details.
}
\end{figure}

Our methodologies are based on order-$N$ (O($N$)) theories,
in which the computational cost is 
proportional to $N$,
as shown in Fig.~\ref{fig-BENCH}(a).  
A fundamental theory of the order-$N$ electronic structure calculation is 
that the electronic structure calculation can be carried out, 
unlike conventional ones, without eigen-value problems. 
\cite{KOHN-1996}
Other references of the order-$N$ calculations are found,
for example, in the literature of  Ref.~\cite{TENG-2011}.
The present calculation was carried out with modeled 
(tight-binding-form) electronic structure theory based on {\it ab initio} calculations. 

The present algorithms are suitable to massive parallelism and 
Fig.~\ref{fig-BENCH}(b) shows 
recent calculations with one-hundred-million atoms on the K computer. 
\cite{HOSHI-mArnoldi, HOSHI-2013-KEI-BENCH}
A one-hundred-million-atom calculation are called  
\lq 100-nm-scale calculation',
because one hundred million atoms are 
those in silicon single crystal with the volume of (126nm)$^3$.

%--------------------------------

%%%%%%%%%%%%%%%%%%%%%%%%%%%%%%%%%%%%%%%%%%
%%%%%%%%%%%%%%%%%%%%%%%%%%%%%%%%%%%%%%%%%%

%%%%%%%%%%%%%%%%%%%%%%%%%%%%%%%%%%%%%%%%%%
\begin{figure}[htbp] 
\begin{center}
  \includegraphics[width=14cm]{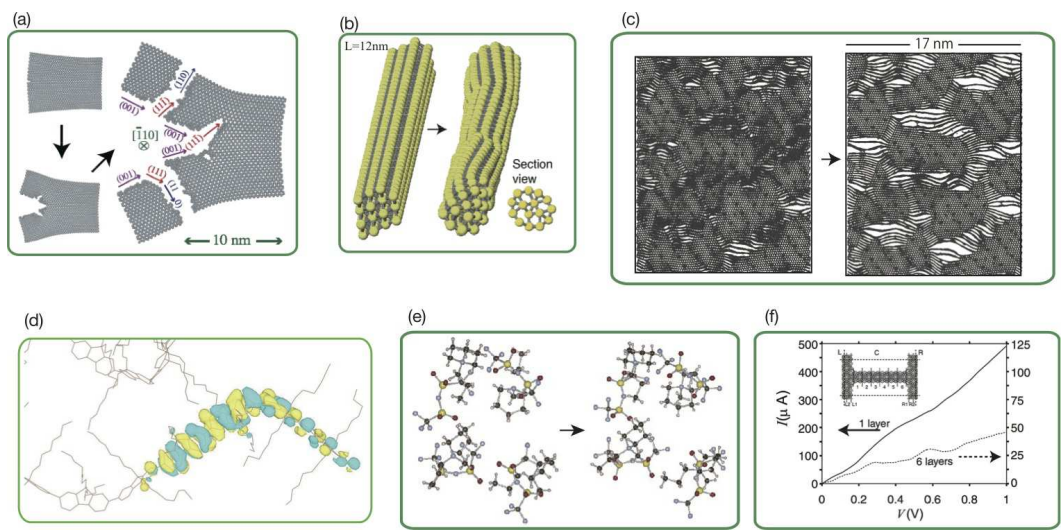}
 %  \vspace{8cm}
\end{center}
%\vspace{-5mm}
\caption{\label{fig-ELSES-appl} 
Examples of nano material study with ELSES;
(a) brittleness of solid silicon, \cite{HOSHI-2005-PRB} 
(b) helical multishell gold nanowire 
\cite{IGUCHI-2007-PRL, HOSHI-2009-JPCM-AUNW, HOSHI-AUNW-HANDBOOK}, 
(c) sp$^2$-sp$^3$ nano-composite carbon solid \cite{HOSHI-2013-KEI-BENCH}, 
(d) amorphous-like conjugated polymer, 
\cite{HOSHI-mArnoldi, HOSHI-2013-KEI-BENCH-APPC}
as a foundations of opto-electronics, 
(e) ionic liquid of 
N-Methyl-N-propylpiperidinium bis trifluoromethanesulfonyl imide
\cite{NISHINO-2012-SSI}, 
as a battery-related problem,
(f) quantum transport of d-band metal nanowire 
\cite{SHINAOKA-2008-NEGF}.
}
\end{figure}
%%%%%%%%%%%%%%%%%%%%%%%%%%%%%%%%%%%%%%%%%%

Figure \ref{fig-ELSES-appl} 
shows examples of nano-material studies with ELSES and 
related original softwares;
(a) brittleness of solid silicon \cite{HOSHI-2005-PRB}, 
(b) helical multishell gold nanowire 
\cite{IGUCHI-2007-PRL, HOSHI-2009-JPCM-AUNW, HOSHI-AUNW-HANDBOOK}
(Experimental work is Ref.~\cite{KONDO-SCIENCE-2000}),
(c) sp$^2$-sp$^3$ nano-composite carbon solid \cite{HOSHI-2013-KEI-BENCH}
in a research of nano-polycrystalline diamond \cite{IRIFUNE-NPD}, 
(d) amorphous-like conjugated polymer, 
\cite{HOSHI-mArnoldi, HOSHI-2013-KEI-BENCH-APPC}
(e) ionic liquid of 
N-Methyl-N-propylpiperidinium bis trifluoromethanesulfonyl imide
\cite{NISHINO-2012-SSI}, 
(f) quantum transport of d-band metal nanowire 
\cite{SHINAOKA-2008-NEGF}.
The studies are motivated by the view points of 
(i)  industrial application, 
(ii) new material, especially new material from Japan, 
and (iii) standard material. 

A Python-based visualization software \lq VisBAR' 
\cite{URL-VISBAR, HOSHI-2013-KEI-BENCH, HOSHI-2013-KEI-BENCH-APPC}
was also developed,
since a post-simulation analysis of large-scale simulations requires 
seamless procedures of large-data-size numerical analysis and visualization.
For example, Fig. \ref{fig-ELSES-appl}(c) was drawn by VisBAR,
when the sp$^2$-sp$^3$ nano-composite carbon solid \cite{HOSHI-2013-KEI-BENCH}
was analyzed 
for the distinction between the sp$^2$ (graphite-like) and sp$^3$ (diamond-like) domains.
The analysis method is a quantum mechanical analysis called $\pi$-type 
Crystalline Orbital Hamiltonian Population ($\pi$COHP).
The method is based on the Green's function for electronic wavefunction
and a theoretical extension of COHP \cite{COHP}.

\section{Novel linear algebraic theory and benchmark on the K computer}

The mathematical foundation of our theories is based on 
the \lq generalized shifted linear equation', 
or the set of  linear equations 
\begin{eqnarray}
 ( z B - A ) \bm{x} = \bm{b},
 \label{EQ-SHIFT-EQ}
\end{eqnarray}
instead of the original generalized eigen-value equation of Eq.~(\ref{EQ-GEV-EQ}).
Here $z$ is a given complex number and has a physical meaning of (complex) energy. 
The vector $\bm{b}$ is the input and the vector $\bm{x}$ is the solution vector. 

Recently, we have developed a set of Krylov subspace (iterative) solvers for Eq.~(\ref{EQ-SHIFT-EQ}); 
(i) generalized shifted conjugate orthogonal conjugate gradient (gsCOCG) method, \cite{TENG-2011} 
(ii) generalized shifted quasi-minimal residual (gsQMR) method, \cite{SOGABE-2012-GSQMR}
(iii) generalized Lanczos (gLanczos) method, \cite{TENG-2011}
(iv) generalized Arnoldi (gArnoldi) method, \cite{TENG-2011}
(v) Arnoldi ($M,W,G$) method, \cite{YAMASHITA-2011}
(vi) multiple Arnoldi method. \cite{HOSHI-mArnoldi}
In the case of $B=I$,
the above theories will be reduced to the previous ones.
\cite{TAKAYAMAI-2004-JPSJ, TAKAYAMA-2006-PRB}
The method is purely mathematical and 
is applicable to other scientific problems. 
For example, one of the above algorithms
was applied to strongly-correlated electrons of La$_{2-x}$Sr$_x$NiO$_4$($x=1/3, 1/2$)
\cite{YAMAGEN-PRG-2007-LASRNIO,YAMAGEN-2008-SCOCG}.

The multiple Arnoldi method \cite{HOSHI-mArnoldi} is mainly used in our simulations
and Fig.~\ref{fig-BENCH} shows the result of benchmark. 
The calculated materials are 
amorphous-like conjugated polymer
of  poly-(9,9 dioctyl-fluorene) (aPF) \cite{HOSHI-mArnoldi, HOSHI-2013-KEI-BENCH-APPC} and
sp$^2$-sp$^3$ nano-composite carbon solid (NCCS).
\cite{HOSHI-2013-KEI-BENCH}.
Figure \ref{fig-BENCH}(a) shows that the calculation 
has the order-$N$ scaling property \cite{HOSHI-mArnoldi}.
Figure \ref{fig-BENCH}(b) shows 
the parallel efficiency on the K computer
with one hundred million atoms. 
The MPI/OpenMP hybrid parallelism is used.
The results are shown for 
a NCCS case with $N=103,219,200$  \cite{HOSHI-2013-KEI-BENCH-APPC} 
and aPF case with $N=102,238,848$ (the present work). 
The elapse time  $T$ of the electronic structure calculation for a given atomic structure 
is measured as the function of 
the number of used processor cores $P$ $(T=T(P))$,
from  $P = P_{\rm min} \equiv 32,768$ to $P_{\rm all} \equiv 663,552$ 
(the total number of processor cores on the K computer). 
The parallel efficiency is defined as $\alpha(P) \equiv T(P)/T(P_{\rm min})$ and
such a benchmark is called \lq strong scaling' in the high-performance computation society. 
The measured parallel efficiency for the NCCS case is
$\alpha(P=98,304)=0.98$, 
$\alpha(P=294,921)=0.90$ and 
$\alpha(P=P_{\rm all})=0.73$. \cite{HOSHI-2013-KEI-BENCH-APPC}  
The measured parallel efficiency for the aPF case is
$\alpha(P=196,608)=0.98$ and 
$\alpha(P=P_{\rm all})=0.56$.

%%%%%%%%%%%%%%%%%%%%%%%%%%%%%%%%%%%%%%%%%%
\begin{figure}[htbp] 
\begin{center}
  \includegraphics[width=8cm]{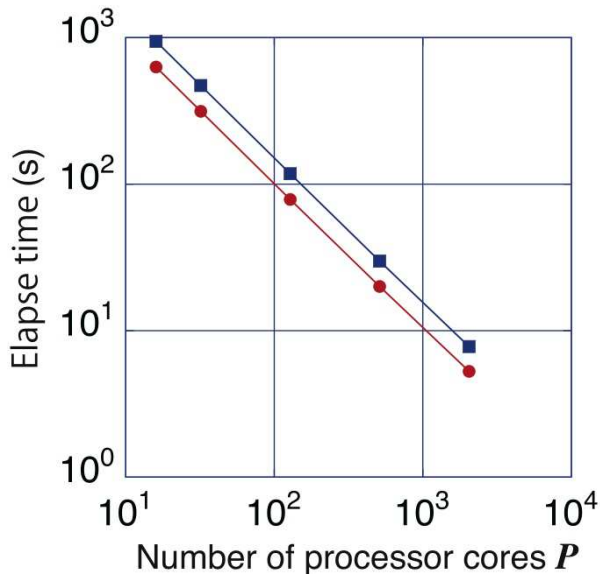}
 %  \vspace{8cm}
\end{center}
%\vspace{-5mm}
\caption{\label{fig-BENCH-FX10} 
Benchmark of the \lq novice'  mode (square) 
and the \lq expert' mode (circle) with $N$=132,864 atoms.
Fujitsu FX10 was used
with $P$ = 16, 32, 128, 512, 2048 processor cores.  
The calculation was carried out with an aPF case. 
}
\end{figure}
%%%%%%%%%%%%%%%%%%%%%%%%%%%%%%%%%%%%%%%%%%

%%%%%%%%%%%%%%%%%%%%%%%%%%%%%%%%%%%%%%%%%%
\section{\lq Novice'  and \lq expert' modes for general users}

A \lq novice' or easy-to-use mode was developed, as well as an \lq expert' mode, 
so as to give comfortable user experiences for everyone.
Nowadays, quantum material simulations are popular among various researchers
and many of them are not familiar to numerical algorithms and detailed procedures of the code.
Such \lq novice' users would like to 
use the code easily in a satisfactory computational performance
with, typically, $10^3$ processor cores or less.
\lq Expert' users, on the other hand,
would like to achieve the best computational performance 
for largest problems with a top-class supercomputer. 
Therefore we developed modes for \lq novice' and \lq expert' users. 

The \lq novice' and \lq expert' modes are different in the computational costs,
though they are equivalent mathematically and give the same numerical results.
The \lq expert' mode is a strict order-$N$ procedure and
requires users to determine the detailed settings for the best performance,
while the \lq novice' mode contains
several $O(N^2)$ procedures
and does not require the detailed settings.
Consequently, 
the \lq novice' mode is easier to use than the \lq expert' mode
but may be worse in the computational performance.

Figure \ref{fig-BENCH-FX10}  demonstrates that
the performance difference
between the two modes is small or moderate 
with a small number of $N$.
Figure \ref{fig-BENCH-FX10} shows a benchmark of $10^5$-atom systems 
with up to $10^3$ processor cores. 
The elapse time of the \lq novice' mode is approximately 50 \% larger than
that of the \lq expert' mode.

%%%%%%%%%%%%%%%%%%%%%%%%%%%%%%%%%%%%%%%%%%
\section{Use of Extensible Markup Language (XML)}

In our simulations, 
the main input and output files are written in the format of Extensible Markup Language (XML),
since the XML format is simple, flexible and widely used on the internet.
For example, 
the minimum information for an atom is written as follows;
\begin{verbatim}
<atom element="C">
  <position unit="angstrom">  1.0d0  0.0d0   0.0d0  </position>
</atom>
\end{verbatim}
The above description means that a carbon atom is located 
at the position of $(x,y,z)=(1, 0, 0)$,
where Angstrom unit (1 Angstrom = 10$^{-10}$m) is used.

The method for reading XML files should be chosen properly,
according to the purpose. 
In our simulations, 
the  XML files are read by two methods, 
Document Object Model (DOM) method and Simple API for XML (SAX) method.
In general, the DOM method is easier in programming and results in huge
memory and time consumption for large-size data,
while the SAX method 
is more difficult in programing and results in tiny memory and time consumption. 

Two input files,  configuration and structure XML files should be prepared for each simulation
and they are quite different in their file size. 
(i) The configuration XML file describes calculation conditions,
such as temperature of the system.
A typical file consists of several tens of lines and 
the file size does not depend on the system size $N$. 
The configuration file is read by the DOM method, since its file size is always tiny.
(ii) The structure XML file describes the atomic structure data, as shown in the above example.
The structure XML file is read by the DOM method, 
since the file size is proportional to the system size $N$ and can be huge. 
Since the atomic structure data
contains three ($x,y,z$) components in the double precision (8B) value for each atom,
the required data size with $10^7$ (=10M) atoms is estimated to be
$3 \times 8 {\rm B} \times 10$M = 240 MB. 
A typical size of the structure XML file with $10^7$ atoms is one G byte (B).
In addition, 
the parallel file reading is used for large-scale calculations 
with split XML files for the structure file and 
gives a significant acceleration.
\cite{HOSHI-2013-KEI-BENCH}
The K computer and FX10 support the parallel file IO,
called \lq rank directory' function, at the hardware level 
and are suitable to the parallel file reading with split XML files. 

%%%%%%%%%%%%%%%%%%%%%%%%%%%%%%%%%%%%%%%%%%
\section{Sparse matrix collection  and numerical \lq engine'}

This section explains that
we opened a matrix library and are developing a general numerical \lq engine',
for further collaboration between real application and numerical linear algebra.

Recently, 
a sparse matrix collection called \lq ELSES matrix library' 
\cite{ELSES-MATRIX-LIBRARY} was opened. 
The stored matrices are sparse and  generated by ELSES 
as the matrices $A$ and $B$ in Eq.~(\ref{EQ-GEV-EQ}).
The maximum matrix size $M$ is one million. 
Matrix data files are written in 
the Matrix Market format. \cite{URL-MATRIX-MARKET}
Each matrix data has its own name and
appears with a document so as to clarify the physical origin of the matrix. 
For example, the matrix data \lq VCNT900' presents
a matrix with the size of $M=900$  
and its physical origin is thermally vibrating carbon nanotube (VCNT).
\footnote{
The thermal vibration causes small random deviations on the atomic positions
and, therefore,  the calculated eigen values are not degenerated. 
}
The calculated eigen values are also included in several matrix data.

Moreover a general numerical 'engine' for matrix (eigen-value) equations is  under development 
with a common interface between real applications and numerical solvers. 
A mini application for evaluating the engine, called \lq Eigen Test', is also under development, 
in which the matrix data in ELSES matrix library \cite{ELSES-MATRIX-LIBRARY}, or 
some files on Matrix Market \cite{URL-MATRIX-MARKET}, can be used as inputs. 
A code in an early version for shared memory systems is available for test users. \cite{EIGEN-TEST}
The code for distributed memory systems is under development, in which two direct solvers, 
ScaLAPACK \cite{URL-SCALAPACK} and EigenExa \cite{URL-EIGEN-EXA}, are implemented 
for the standard and generalized eigen-value equations in the Eq.~(\ref{EQ-GEV-EQ}) with 
real-symmetric matrices.
\footnote{
At the present time, 
EigenExa does not support generalized eigen-value problems.
In the engine, a generalized eigen-value problem is solved with EigenExa, 
since the problem is transformed into a standard eigen-value problem
by the Cholesky decomposition routine in ScaLAPACK.
}
The implementation of Krylov subspace solvers is planned. 
In near future,
our application code \lq ELSES' will be connected to the engine
and general users can use the solvers without detailed knowledge of the solver algorithms.
The engine is general and can be connected to any other real applications.

%%%%%%%%%%%%%%%%%%%%%%%%%%%%%%%%%%%%%%%%%%
\begin{figure}[htbp] 
\begin{center}
  \includegraphics[width=12cm]{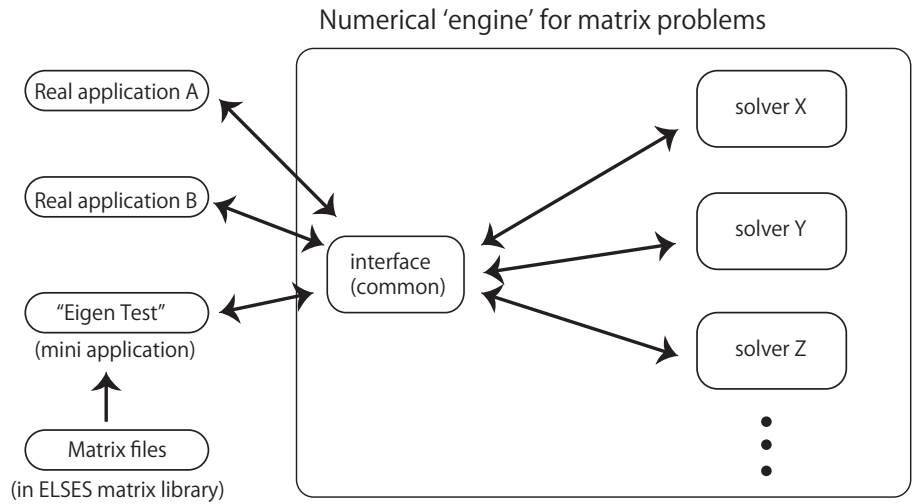}
 %  \vspace{8cm}
\end{center}
\caption{\label{fig-eigen-engine} 
Illustration of the role of the general \lq numerical engine' for matrix equations. 
}
\end{figure}

%%%%%%%%%%%%%%%%%%%%%%%%%%%%%%%%%%%%%%%%%%
\section{Internal eigen-value problem}

%%%%%%%%%%%%%%%%%%%%%%%%%%%%%%%%%%%%%%%%%%
\begin{figure}[htbp] 
\begin{center}
  \includegraphics[width=12cm]{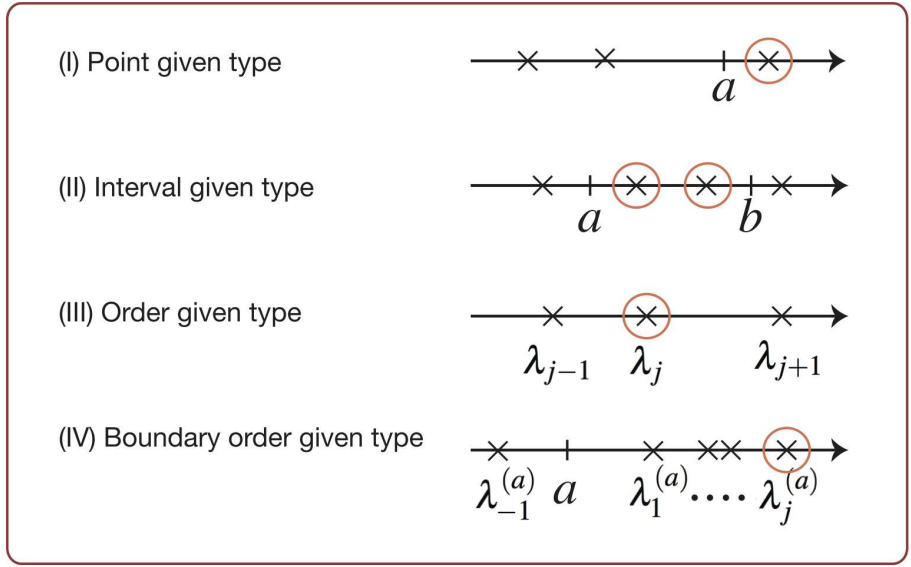}
 %  \vspace{8cm}
\end{center}
%\vspace{-5mm}
\caption{\label{FIG-INTERNAL-EIGEN} 
Four types of internal eigen-value problems.
(I) Point given type,
(II) Interval given type, 
(III) Order given type,
(IV) Boundary order given type.
}
\end{figure}
%%%%%%%%%%%%%%%%%%%%%%%%%%%%%%%%%%%%%%%%%%

This section explains internal eigen-value problem
from a general need in large-scale electronic state calculations.
The present discussion is limited to real eigen-value problems.
In a large-matrix problem, 
one should give up the calculation of all the eigen pairs, 
because of huge computational cost.
Then one would like to obtain specific eigen pair(s)  
($\lambda_j, \bm{y}_j$) of Eq.~(\ref{EQ-GEV-EQ}) 
at an internal eigen-value region ($1 < j < M$),
because 
several internal eigen pairs are responsible 
for electronic and optical properties.

Internal eigen-value problems can be classified 
into the following four types, 
as illustrated in Fig.~\ref{FIG-INTERNAL-EIGEN};
(I) Point given type:
For a given real value $a$,
one should find (an) eigen value(s) $\lambda$ near the given point
($\lambda \approx a$) and its (their) eigen vector(s). 
(II) Interval given type:
For a given interval of $[a,b]$  ($a < b$),
one should find eigen values $\lambda$ in the interval  
($a < \lambda < b$) and their eigen vectors. 
(III) Order given type:
For a given integer of $j$ ($ 1 < j < M$),
one should find the $j$-th lowest eigen value 
$\lambda_j$ ($\cdot \cdot \cdot \lambda_{j-1} \le \lambda_j \le \lambda_{j+1} \cdot \cdot \cdot$) and its eigen vector. 
(IV) Boundary order given type:
For a given integer  $j$ ($ 1 < j$) and a real value $a$,
one should find the $j$-th lowest eigen value $\lambda_j^{(a)}$
that is larger than $a$:
\begin{eqnarray}
\cdot \cdot \cdot \le \lambda_{-1}^{(a)} \le a \le \lambda_1^{(a)} \le \lambda_2^{(a)} \le \cdot \cdot \cdot \lambda_j^{(a)} \le \lambda_{j+1}^{(a)} \cdot \cdot \cdot
\end{eqnarray}
One should find all the (degenerated) eigen vectors
for the target eigen value $\lambda_j^{(a)}$.
From the definitions, 
the boundary-order-given-type problem with $a < \lambda_1$
is reduced to the order-given-type problem.
In all the types of problem, 
if the target eigen value is degenerated, one should find all the degenerated eigen vectors.

The present paper focuses on the order-given-type problem,
since the problem appears in many electronic structure calculations.
In our problems, 
the target eigen value(s) is (are) given from the physical viewpoint.
Here we call eigen value \lq level', as usual in electronic structure calculations.
Typically, the two levels, called highest-occupied (HO) and lowest-unoccupied (LU) levels,
are of fundamental interest.  
The HO level, denoted as $j$ hereafter, is defined as
the half of the total number of electrons in material $N_{\rm e}$
($j \equiv [ N_{\rm e} / 2 ]$)
\footnote{The present case is that of a para-spin material.}
and the LU level is defined as $j$+1. 
For example,
the difference between the two levels
$\Delta \equiv \lambda_{j+1} -  \lambda_{j}$
is called energy gap
and is zero in metals and non-zero in semiconductors or insulators.
Two notes are added here;
(a) The levels near the HO or LU level are also of importance 
and are required in many electronic state calculations. 
(b) When $N$ is odd,
the HO and LU levels are not defined in a strict physical sense.
Among these cases, however,
the ($j \equiv [ N_{\rm e} / 2 ]$])-th  and ($j$+1)-th  levels are still crucial
for electronic properties and
we call them the HO and LU levels in the present paper.

We have proposed two approaches for the order-given-type problem:
One appears in Refs.~\cite{LEE-2013-INTERIOR-EIGEN, LEE-2013-EASIAM}
and the other appears in Ref.~\cite{HOSHI-2013-KEI-BENCH-APPC}.  
See the papers for details.

Further investigations on efficient algorithms are desired
for internal eigen-value problems.
It is speculated that a difficulty in numerical treatment 
will appear among (almost) degenerated eigen pairs.
An example is found 
in the matrix of \lq APF4686'
of ELSES matrix library
\cite{ELSES-MATRIX-LIBRARY},
in which the 2345-th and 2346-th eigen values 
are almost degenerated 
($\lambda_{2345}$=-0.356883, $\lambda_{2346}$= -0.356806).
\footnote{The atomic unit is used in the present eigen values}

%%%%%%%%%%%%%%%%%%%%%%%%%%%%%%%%%%%%%%%%%%
\begin{figure}[htbp] 
\begin{center}
  \includegraphics[width=14cm]{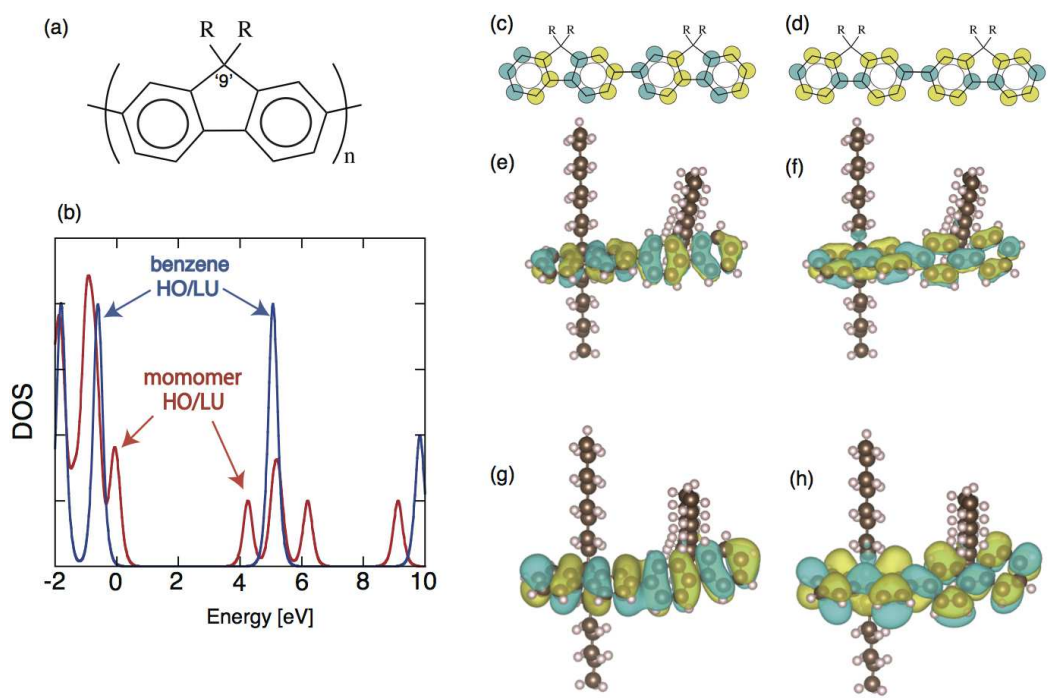}
 %  \vspace{8cm}
\end{center}
\caption{\label{fig-PF-monomer-dimer} 
(a) Structure of poly-(9,9 dioctyl-fluorene).  
Here $R \equiv$ C$_8$H$_{17}$ and $n$ is the number of monomer units.
The letter \lq 9' indicates the atom site called \lq 9 position'. 
(b) DOS of the monomer (red line) and benzene (blue line).
The value of the latter graph is magnified twice. 
(c)(d)  Schematic figures of
the HO or LU wavefunction  of the dimer, respectively.  
(e)(f) The calculated dimer structure and the HO or LU wavefunctions, respectively.
The large and small balls indicate carbon and hydrogen atoms. 
The colors indicates the sign of the wavefunction in (c)-(h). 
}
\end{figure}

%%%%%%%%%%%%%%%%%%%%%%%%%%%%%%%%%%%%%%%%%%
\section{Physical aspect of the fundamental methodologies}

The last topic is some physical (not mathematical) aspects of the fundamental methodology,
in particular, {\it ab initio} based modelings in electronic structure theory.
The present calculation was carried out with modeled 
(tight-binding-form) electronic structure theory based on {\it ab initio} calculations and 
sometimes a charge-self-consistent theory \cite{ELSTNER-1998} is also used 
\cite{NISHINO-2012-SSI, NISHINO-2013-JMODELING}.   
The matrices $A$ and $B$ 
in Eqs.(\ref{EQ-GEV-EQ}) and (\ref{EQ-SHIFT-EQ})
are sparse and called 
the Hamiltonian and overlap matrices 
written the linear-combination-of-atomic-orbital (LCAO) representation, respectively.
The diagonal elements of $B$ are the unity $(B_{ii}=1)$ and 
the absolute value of an off-diagonal element is less than the unity
$(| B_{ij} |  < 1, i \ne j)$.

Here  poly-(9,9 dioctyl-fluorene), 
a conjugated polymer depicted in Fig.~\ref{fig-PF-monomer-dimer}(a),
is picked out.
The material is a basic one for opto-electronics devices.
A modeled theory in Ref~\cite{ASED-CALZAFFERI} is used
and it is called the atomic-superposition and electron-delocalization 
tight-binding (ASED-TB) theory.
The monomer unit in Fig.~\ref{fig-PF-monomer-dimer}(a)
consists of two benzene rings 
and alkyl side chains of $R \equiv$ C$_8$H$_{17}$.  
The letter \lq 9' indicates the atom site called \lq 9-position',
located at the \lq root' part of the side chain or
the top site of the five-membered ring.
The side chains are extended transversely,
since the \lq 9-position' in Fig.~\ref{fig-PF-monomer-dimer}(a)
has an $sp^3$-like electronic configuration and
the neighboring four atoms form tetrahedral bonds.

The monomer and dimer were calculated with the exact solution of Eq.~(\ref{EQ-GEV-EQ}).
The density of states (DOS) function, a eigen-value histogram, 
is shown in Fig.~\ref{fig-PF-monomer-dimer}(b) for the  monomer and a benzene molecule (C$_6$H$_6$). 
The HO and LU electronic wavefunctions of the dimer are drawn schematically
in Fig.~\ref{fig-PF-monomer-dimer}(c)(d), respectively
and the calculated HO and LU wavefunctions with the optimized structure
are shown in Fig.~\ref{fig-PF-monomer-dimer}(e)(f), respectively.
The two monomer units of the dimer
are connected with twisting, as shown in Fig.~\ref{fig-PF-monomer-dimer}(e) or (f),  
mainly because of the repulsion 
between the hydrogen atoms on the benzene rings of the neighboring monomers.
The half of Fig.~\ref{fig-PF-monomer-dimer}(c) or (d) 
corresponds to the schematic figure of 
the HO or LU state of the monomer, \cite{QM-TEXT} respectively. 
The wavefunctions are contributed only by  the $\pi$-type wavefunctions on the benzene rings.
The HO wavefunction of the dimer and the monomer has  two nodes on a benzene ring, 
and the LU wavefunction of the dimer and the monomer has four nodes on a benzene ring.
These node structures are the same with those of a benzene,
\cite{QM-TEXT}
because the HO and LU levels of the monomer stem from those of benzene,
which is seen in the DOS functions of Fig.~\ref{fig-PF-monomer-dimer}(b).
The monomer and dimer are calculated also 
by the {\it ab initio} calculation (Gaussian 09$^{\rm (TM)}$)
with the B3LYP functional and  the 6-31G(d,p) basis set
and the above features of wavefunctions are satisfied 
in the calculated HO and LU wavefunctions shown in Figs.~\ref{fig-PF-monomer-dimer}(g)(h).
The atomic structures and the DOS functions were also calculated
and satisfy the above features. 
Moreover, detailed data by the present method are added here
with those by the {\it ab initio} calculation in the parentheses; \cite{HOSHI-mArnoldi} 
The balance band width $W$ and the band gap $\Delta$ are
$W=18.5$ eV (18.3 eV) and $\Delta =4.25$ eV (4.91 eV) in the monomer
and $W=19.0$ eV (18.8 eV)  and $\Delta = 3.58$ eV (4.10 eV) in the dimer.
The twisting angle of the dimer $\theta$ is 
$\theta=37.3 ^\circ$ (40.6 $ ^\circ$).

Several notes are posted on the transferability (general applicability) of model theories;
(i) The present theory is formulated without any data of fluorene cases but
is formulated with related small molecules such as benzene.
The above agreement with the {\it ab initio} calculations in the fluorene cases shows
that the theory is applicable to, at least, materials that have similar binding mechanisms.
(ii) Quite recently, a modeled van der Waals (vdW) interaction 
\cite{ORTMANN-2006} is implemented in the code 
for wider applicability to  materials. 
Details will be discussed elsewhere.
(iii) Automated determination methods for obtaining an optimal model among various materials is desired and
is now developing with ELSES \cite{NISHINO-2013-JMODELING, OHTANI-2013-MRS}.

%%%%%%%%%%%%%%%%%%%%%%%%%%%%%%%%%%%%%%%%%%

%Acknowledgments}

\section{Summary and future outlook}

The present paper reports our methods and results of 
large-scale electronic structure calculations
based on our novel linear algebraic algorithms. 
A high parallel efficiency was shown 
in one-hundred-million-atom systems 
with up to all the built-in cores of the K computer. 
The related methodologies on physics, mathematics and information technology are also discussed. 

The present study is a interdisciplinary one between 
physics, numerical linear algebra and high-performance computation techniques,
and such a interdisciplinary study is sometimes called 'Application-Algorithm-Architecture co-design'.
The co-design will play a crucial role in exa-scale scientific computations.

%%%%%%%%%%%%%%%%%%%%%%%%%%%%%%%%%%%%%%%%%%
\acknowledgments
This research is partially supported by Grant-in-Aid 
for Scientific Research
(KAKENHI Nos. 23540370 and 25104718)
from the Ministry of Education, Culture, Sports, Science and Technology 
(MEXT) of Japan. 
The K computer was used 
in the research proposals of hp120170, hp120280 and hp130052.
Several calculations were carried out 
by the supercomputer
at the Information Technology Center, University of Tokyo,
in the research proposal of jh130011-NA07.
Supercomputers were also used 
at the Institute for Solid State Physics, University of Tokyo, 
and at the Research Center for 
Computational Science, Okazaki.

\end{document}